\documentclass[dvips]{article}

\usepackage{icrc2011}

\title{Emission models and EBL as a tool to measure the redshift of BL Lac objects}
\newcommand{\etal}{\MakeLowercase{\textit{et al. }Redshift estimation from emission model and EBL}} % "et al."
\shorttitle{Mankzuhiyil \etal }

\authors{Nijil Mankuzhiyil$^{1}$, Stefano Ansoldi$^{2}$, Gessica De Caneva$^{3}$, Massimo Persic$^{4}$, Fabrizio Tavecchio$^{5}$  }
\afiliations{$^1$University of Udine \& INFN Trieste, Italy\\
$^2$ ICRA Roma \& INFN Trieste \& University of Udine, Italy\\
$^3$ DESY Zeuthen, Germany \\
$^4$ INAF \& INFN Trieste, Italy\\
$^5$ INAF Milano, Italy}
\email{nijil.mankuzhiyil@uniud.it}

\abstract{

We introduce a new method to determine the redshift of unknown-redshift BL Lac
Objects. The method relies on simultaneous multi-wavelength (MWL) observations of BL
Lac objects in optical, X-ray, HE (E>100 MeV) gamma-rays and VHE (E>100 GeV)
gamma-rays. It involves best-fitting spectral energy distribution (SED) from optical
through HE gamma-rays with a Synchrotron-Self-Compton (SSC) model. We extrapolate
such best fitting model into VHE regime, and assume that it represents the intrinsic
emission of the object. We then compare the observed VHE flux which has been affected
by the interaction with Extragalactic Background Light (EBL). Constraining the
measured vs intrinsic emission leads to the determination of gamma-gamma opacity.
Comparing the obtained opacity with the predicted opacity based on EBL model, we
obtain the redshift of the photon source.

}

\begin{document}
\maketitle

%Begin the section.
\section{Introduction}

%\subsection{File format}

In a recent paper \cite{bib:man010}, some of us (NM, MP and VT) have shown how it is possible to use
simultaneous multiwavelength (i.e. optical, X-ray, HE\,$\gamma$, and VHE\,$\gamma$) data to obtain a
measure the Extragalactic Background Light (EBL). EBL is the integrated light produced and re-processed
in all the history of the universe. Knowledge of the EBL is particularly important in studying the
VHE\,$\gamma$-ray emission from extragalactic sources (typically, blazars), because interaction of TeV
photons with the EBL attenuates their observed intensity while the resulting cascade will enhance the
observed GeV emission. These effects are clearly more marked for more distant sources, because of the
larger optical depth, $\tau_{\gamma \gamma}$, for the $\gamma_{\rm VHE}-\gamma_{\rm EBL}$ interaction:
for a given EBL model, the quantitative details of the modification to the VHE\,$\gamma$-ray spectrum
emitted by a source will only depend on the redshift of the source.

By reversing the line of reasoning in \cite{bib:man010}, if we assume to have a reliable EBL model, then
simultaneous multi-$\lambda$ observations of a source may allow for an independent derivation of the source
redshift, $z$. For instance: (1) low- and high-energy (eV, keV, GeV) spectral-energy-distribution (SED) data (that are
unaffected by EBL) are fitted with an emission model; (2) the model SED is extrapolated into the TeV domain,
and the extrapolation is assumed to represent the insinsic (emitted) TeV emission; (3) the 'emitted' VHE
spectrum is compared with the measured VHE spectrum, and the discrepancy between the two is taken to be a
measure of $\tau_{\gamma \gamma}$; (4) given a specific EBL model, $\tau_{\gamma \gamma}$ can be solved as
a function of the source distance (i.e., for a given cosmology, of $z$).

In what follows we present preliminary results of an approach to the problem of distance determination for
a blazar, which is slightly different from the one above and in which we employ a dynamical EBL correction
during a SED fit in which $z$ also appears as a parameter. Just below, we describe the source that we are
using for our analysis. Then we provide some details about the numerical algorithm that, elaborating what
is done in \cite{bib:man011}, is used to perform the numerical fit to a chosen SED model which will return,
{\rm inter alia}, an estimation of the source redshift.

\section{The source and the data}

In this preliminary study we choose a source whose redshift is actually known. By independently determining
$z$ with the method detailed below, and comparing the result with the known value, we will gather an indication
on the reliability of our method. The source we consider here is PKS\,2155-304, whose redshift is $z = 0.12$.
We apply our method to the simultaneous data presented in \cite{bib:aha009} and shown in fig.\,1.

 \begin{figure}
  %\vspace{-0.5cm}
\hspace{-1.cm}
  %\centering
  \includegraphics[width=3.2 in]{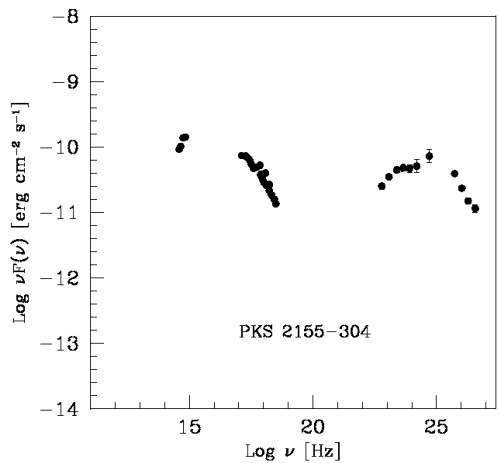}
   \vspace{-1.5cm}
\hspace{-1cm}
  \caption{The dataset of PKS\,2155-304 \cite{bib:aha009}.}
  \label{fig:dataset}
 \end{figure}

\section{The method}

We assume that the description of the emission of PKS\,2155-304 can be understood within a one zone
Synchrotron-Self-Compton (SSC) model: the two peaks, synchrotron and Compton, characteristic of these
models are here understood within the framework detailed in \cite{bib:tav998}. Although different models
have also been employed to describe the emission of PKS\,2155-304, as discussed in \cite{bib:man010} the
model that we are choosing can also be used to fit the data and does not provide quantitatively significant
discrepancies in the characteristic parameters that describe the source emission; in particular the fit of
the data to the \cite{bib:tav998} faithfull SED can be done as described in \cite{bib:man011} using the
Levenberg--Marquardt method \cite{bib:numrec}, which is an efficient standard for non-linear least-squares
minimization that smoothly interpolates between the inverse Hessian method and the steepest descent method.

In \cite{bib:man011} the known redshift of the source was used to correct VHE data, before performing the fit.
In the present analysis, instead, $z$ is what we wish to determine. We can obtain our result by a refinement
of the fit algorithm described in \cite{bib:man011}, in which $z$ is added as a parameter to be estimated by
the fit procedure. Of course, while the $\chi ^{2}$ minimization procedures explores the parameter space,
different $z$ values will also be explored. At each step in which $z$ changes, we need to apply an EBL
correction appropriate for the $z$ that is being considered. In other words, EBL correction has to be performed
dynamically during the fit. The EBL model we have chosen is the one by \cite{bib:fra008}, which represents the
minimum EBL level because it is based on measured galaxy counts. EBL corrections have been added to the code
for several possible values of $z$. Linear interpolation is used to determine the value of the EBL correction
at a given energy if the $z$ that is currently used in the $\chi^2$-minimization procedure falls within to
tabulated values. The flow-chart of the code that we employed is shown in fig.\,2.

 \begin{figure}
  \vspace{5mm}
  %\centering
  \hspace{-1.cm}
  \includegraphics[width=4.in,height=2.3 in]{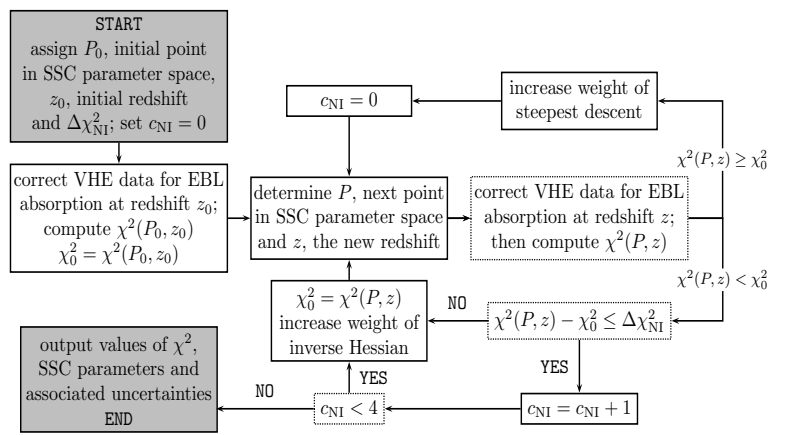}
\vspace{0.5cm}
  \caption{The flow chart of the code}
  \label{simp_fig}
 \end{figure}

\section{Preliminary results}

Our results show that the procedure described above is able to fit the SED parameters and the redshift
of PKS\,2155-304, and that all the derived values are compatible with the published values for the source
parameters.

The statistical analysis of the significance of the fit is currently underway. Preliminary results suggest
that blazar SED fitting to simultaneous multi-$\lambda$ data with dynamical EBL correction can indeed be used
to provide redshift estimations for these sources. A more in-depth analysis will be presented elsewhere
\cite{bib:wip011}. It will also be important to check the stability of the results versus the choice of both
the emission model and the EBL model -- especially in cases where no other methods for redshift estimation can
be used.

\clearpage

\end{document}